\documentclass[ja4paper]{IEEEtran}

\usepackage[utf8]{inputenc} 
\usepackage[T1]{fontenc}
\usepackage{url}              % provides \url{...}
\usepackage{cite}             % improves presentation of citations

\usepackage[cmex10]{amsmath}  % Use the [cmex10] option to ensure complicance
\usepackage{mleftright}       % fix to wrong spacing of \left-,
\mleftright                   % \middle- \right-commands 

\usepackage{graphicx}         % provides \includegraphics{...} to
\usepackage{booktabs}         % fixes poor spacing in tables and
\usepackage[utf8]{inputenc} % allow utf-8 input
\usepackage[T1]{fontenc}    % use 8-bit T1 fonts
\usepackage{url}            % simple URL typesetting
\usepackage{booktabs}       % professional-quality tables
\usepackage{amsfonts}       % blackboard math symbols
\usepackage{nicefrac}       % compact symbols for 1/2, etc.
\usepackage{microtype}      % microtypography
\usepackage{xcolor}         % colors
\usepackage{amsmath, amssymb}      % microtypography
\usepackage{enumerate}
\usepackage{cuted}
\usepackage{lipsum}
\usepackage{balance}
\usepackage{nicematrix}
\usepackage{bigstrut}

\usepackage{tikz} % To generate the plot from csv
\usepackage{pgfplots}
\usepackage{pgfplotstable}
\usetikzlibrary{patterns}
\pgfplotsset{compat = newest}
\pgfkeys{/pgf/number format/set thousands separator = }
\usepackage{filecontents}
\usetikzlibrary{decorations.pathmorphing, graphs,graphs.standard} 
\usetikzlibrary{automata}
\usepgfplotslibrary{statistics}
\usetikzlibrary{arrows,chains,matrix,positioning,scopes}
\usetikzlibrary{shapes.geometric}
\pgfplotsset{compat = newest}
\pgfkeys{/pgf/number format/set thousands separator = }
\definecolor{chocolate1}{rgb}{0.94,0.7,0.5}

\newcommand{\R}{\mathbb{R}}
\newcommand{\C}{{\mathbb C}}

\newcommand{\K}{\mathcal{K}}

\newcommand{\G}{\mathcal{G}}
\newcommand{\V}{\mathcal{V}}
\newcommand{\U}{\mathcal{U}}
\newcommand{\W}{\mathcal{W}}
\newcommand{\E}{\mathcal{E}}

\newcommand{\X}{\mathbf{X}}
\newcommand{\x}{{\mathbf x}}
\newcommand{\y}{{\mathbf y}}
\newcommand{\z}{{\mathbf z}}

\newcounter{examplecntr}
{\begin{trivlist}\small\item[]\refstepcounter{examplecntr}%
 {\bfseries Example~\theexamplecntr%
  \ifthenelse{\equal{#1}{}}{}{ (#1)}.
}}%
{\end{trivlist}}

\newcounter{propositioncntr}
{\begin{trivlist}\item[]\refstepcounter{propositioncntr}%
{\bfseries Proposition~\thepropositioncntr%
  \ifthenelse{\equal{#1}{}}{}{ (#1)}.
}}%
{\end{trivlist}}

\newcounter{remarkcntr}
{\begin{trivlist}\item[]\refstepcounter{remarkcntr}%
{\bfseries Remark~\theremarkcntr%
  \ifthenelse{\equal{#1}{}}{}{ (#1)}.
}}%
{\end{trivlist}}

\newcounter{theoremcntr}
{\begin{trivlist}\item[]\refstepcounter{theoremcntr}%
{\bfseries Theorem~\thetheoremcntr%
  \ifthenelse{\equal{#1}{}}{}{ (#1)}.
}}%
{\hfill$\Box$\end{trivlist}}

\begin{document}

\title{On the Equivalence of Gaussian Graphical Models\\Defined on Complete Bipartite Graphs} 

\author{\IEEEauthorblockN{Mehdi Molkaraie}\\
\IEEEauthorblockA{Department of Statistical Sciences\\
University of Toronto \\
{\tt mehdi.molkaraie@alumni.ethz.ch}}
}

\maketitle

\begin{abstract}
This paper introduces two Gaussian graphical models defined on complete bipartite graphs. We show 
that the determinants of the precision matrices associated with the models are equal
up to scale, where the scale factor only depends on model parameters.
In this context, we will introduce a notion of ``equivalence" between 
the two Gaussian graphical models. This equivalence has two key applications: first, it can significantly 
reduce the complexity of computing the exact value of the determinant, and 
second, it enables the derivation of closed-form expressions 
for the determinants in certain special cases.

%For a specific class of sparse Gaussian graphical models, we provide a closed-form 
%solution for the determinant of the covariance matrix. In our framework, 
%the graphical interaction model (i.e., the covariance selection model) is equal to replacement
%product of $\mathcal{K}_{n}$ and $\mathcal{K}_{n-1}$, where $\mathcal{K}_n$ is the complete graph with $n$ vertices. 
%Our analysis is based on
%taking the Fourier transform of the local factors of the model, which can be
%viewed as an application of the Normal Factor Graph Duality Theorem and holographic algorithms. 
%The closed-form expression is obtained by applying the Matrix Determinant Lemma on the 
%transformed graphical model. 
%In this context, we will also define a notion of equivalence between 
%two Gaussian graphical models. 
\end{abstract}

\section{INTRODUCTION}
\label{sec:intro}

In this paper, we are concerned with computing the determinant of structured real symmetric 
positive-definite matrices. The determinant of covariance/precision matrices is a crucial quantity in many areas 
of statistics and information theory. Applications include computing the generalized 
variance~\cite{wilks1932certain, Sengupta2004}, the differential 
entropy~\cite{ahmed1989entropy, misra2005estimation}, and 
the \mbox{Kullback–Leibler} divergence~\cite{cover1991elements}, which are important in statistical inference, hypothesis testing, and
classification~\cite{witten2009covariance}.

In general, the determinant of a $p\times p$ (covariance) matrix can be calculated with computational complexity $\mathcal{O}(p^3)$, which is infeasible for 
large $p$. To address this challenge,
lower and upper bounds for the determinant of sparse positive definite matrices are provided in~\cite{bai1996bounds}.
In~\cite{bai1996some, barry1999monte}, Monte Carlo methods have been proposed to estimate the determinant (or the log determinant) of 
sparse (positive definite) matrices. Randomized algorithms have been introduced to approximate the log determinant of positive definite 
matrices (see, e.g., \cite{boutsidis2017randomized}). Further techniques to approximate the determinant of (sparse) positive definite matrices have 
been suggested, including sparse approximate inverses~\cite{reusken2002approximation}, Chebyshev polynomial expansions \cite{han2015large}, 
and adaptive thresholding estimators~\cite{cai2011adaptive}.

We focus on two Gaussian graphical models defined
on complete bipartite graphs. In the first model, the precision matrix is a block matrix, in which the diagonal blocks are complete matrices (i.e., all their entries are nonzero), 
and the off-diagonal blocks are diagonal.
In contrast, in the second model, the diagonal blocks of the precision matrix are diagonal matrices, and the off-diagonal blocks are complete.
The size of the precision matrix of the first model is $n^2\times n^2$, whereas in the second model, the precision matrix is of size $n\times n$.

We will show that the models are dual to each other, and as a result, the determinants of their precision matrices are
equal up to a scale factor. We refer to these models as \emph{equivalent}.

This equivalence has two key benefits: i) by reducing the dimensionality of the problem, it simplifies the computation (or approximation) of the determinant
ii) it enables the derivation of closed-form expressions for the determinant in specific cases, such as homogeneous models.

The paper is organized as follows. Some notation and preliminaries are introduced in Section~\ref{sec:notations}. 
The general structure of the precision matrices is described in Section~\ref{sec:matrices}.
The models and their normal factor graph representations on complete bipartite graphs are 
presented in Section~\ref{sec:NFGEquiv}. Section~\ref{sec:Equiv} established the equivalence by showing that the determinants 
of the two Gaussian models are equal up to scale. 
The exact determinants for homogeneous models and for the special case of the star graph are derived in 
Sections~\ref{sec:Exact} and~\ref{sec:NFGK1N}. 

\section{Notation and preliminaries}
\label{sec:notations}

In this section, we introduce the notation and preliminaries that will be used throughout this paper.

A zero-mean real random vector ${\X}_{p\times 1}$ has a $p$-variate Gaussian distribution if it 
has the following PDF
\begin{equation}
\label{eqn:GaussPDF}
p(\x) =  \frac{1}{|2\pi \pmb{\Sigma}|^{1/2}}
\textrm{exp}\big(-\frac{\ 1\ }{ \ 2\ }
\x_{\phantom{1}}^\intercal
\pmb{\Sigma}^{-1}\x\,\big), \quad \x \in \R_{\phantom{1}}^p
\end{equation}
where $\x_{\phantom{1}}^\intercal = (x_1, x_2, \ldots, x_p)$,
the symmetric positive definite matrix $\pmb{\Sigma}^{-1} \in \R^{p\times p}$ is the 
precision (information) matrix, and $\pmb{\Sigma}$ is the corresponding covariance matrix. 

The structure of a Gaussian graphical model is completely determined by its precision matrix. A
nonzero entry of $\pmb{\Sigma}^{-1}$ indicates the presence of a factor in the graphical model and 
an off-diagonal zero entry of $\pmb{\Sigma}^{-1}$ indicates the lack of pairwise interaction between
the corresponding random variables~\cite{lauritzen1996graphical}, \cite[Chapter 19]{murphy2012machine}. 

All vectors are represented as column vectors.
The all-ones matrix of 
size $p\times p$ is denoted by $\mathbf{J}_{p}$ and the identity matrix of size 
$p\times p$ is denoted by $\mathbf{I}_{p}$. The set of positive integers from 1 to $n$ is denoted by $[n]$.

We focus on complete 
bipartite graphs $\K_{m,n} = (\V ,\E)$, in which the vertex set $\V$ can be partitioned into two 
disjoint subsets  
$\U$ and $\W$ so that each 
edge $e \in \E$ connects a vertex $u \in \U$ to a vertex $w \in \W$. In $\K_{m,n}$, each 
vertex in $\U$ has degree $m$ and
each vertex in $\W$ has degree $n$. Thus $|\V| = m+n$ and $|\E| = mn$.

We use graphical models defined in terms of normal factor graphs. In normal factor graphs variables are 
represented by edges and factors by vertices. Moreover, normal factor graphs allow for a 
simple and elegant graph dualization procedure~\cite{Forney:01}. 

For a subset $\mathcal{T} \subset [n]$, let $\x_\mathcal{T} = (x_t, t \in \mathcal{T})$, and define the 
zero-sum indicator function as
\begin{equation} 
\label{eqn:Definition2}
\delta_{+}(\x_\mathcal{T}) =  \left\{ \begin{array}{ll}
    1, & \text{if $x_1 + x_2 + \ldots + x_{|\mathcal{T}|} = 0$} \\
    0, & \text{otherwise,}
  \end{array} \right.
\end{equation}
and the equality indicator function as
\begin{equation} 
\label{eqn:Definition1}
\delta_{=}(\x_\mathcal{T}) =  \left\{ \begin{array}{ll}
    1, & \text{if $x_1 = x_2 = \ldots = x_{|\mathcal{T}|}$} \\
    0, & \text{otherwise.}
  \end{array} \right.
\end{equation}
Both functions (\ref{eqn:Definition2}) and (\ref{eqn:Definition1}) are equivalent to the dirac 
delta function $\delta(\cdot)$ when $|\mathcal{T}|=1$. 
%Notice that $\delta_{=}(x_k, x_\ell)$ and $\delta_{+}(x_k, x_\ell)$ are both equal to $\delta(x_k - x_\ell)$.

The Fourier transform of a function $f(\x) \colon \R_{\phantom{1}}^p \rightarrow \C$ is the function
$\tilde{f}(\tilde{\x}) \colon \R_{\phantom{1}}^p \rightarrow \C$ given by
\begin{equation}
\label{eqn:ft}
\tilde{f}(\tilde{\x}) = \int_{-\infty}^{\infty} f(\x)e^{-\mathrm{i}\,\tilde{\x}_{\phantom{1}}^\intercal \x}d\x
\end{equation}
where $\mathrm{i} = \sqrt{-1}$ and $\C$ denotes the set of 
complex numbers. In particular, the Fourier transform of the PDF in~(\ref{eqn:GaussPDF}) is
\begin{equation}
\label{eqn:ftGaussian}
\tilde{p}(\tilde{\x}) = \textrm{exp}\big(-\frac{1}{2}
\,\tilde{\x}_{\phantom{1}}^\intercal
\pmb{\Sigma} 
\,\tilde{\x}\big), \quad \tilde{\x} \in \R_{\phantom{1}}^p
\end{equation}
see~\cite{anderson1958introduction}.

\section{The Structure of Matrices}
\label{sec:matrices}

We consider two Gaussian graphical models whose precision matrices are real symmetric 
positive-definite block 
matrices with the following structure:

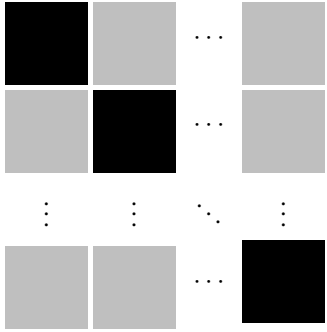
\begin{center}
\begin{tikzpicture}
\matrix [matrix of nodes,column sep=0.4ex,row sep=0.4ex]
  {
   \node[fill=black,rectangle,minimum size=1.1cm]{$$}; & \node[fill=black!25,rectangle,minimum size=1.1cm]{$$}; & $\,\cdots\,$ & \node[fill=black!25,rectangle,minimum size=1.1cm]{$$}; \\
    \node[fill=black!25,rectangle,minimum size=1.1cm]{$$}; &  \node[fill=black,rectangle,,minimum size=1.1cm]{$$}; & $\,\cdots\,$ &  \node[fill=black!25,rectangle,,minimum size=1.1cm]{$$};\\
    $\vdots$ & $\vdots$ &  $\ddots$ & $\vdots$ \\
    \node[fill=black!25,rectangle,,minimum size=1.1cm]{$$}; & \node[fill=black!25,rectangle,minimum size=1.1cm]{$$};&  $\,\cdots\,$ & 
\node[ fill=black,rectangle,minimum size=1.1cm]{$+$}; \\
  };
\end{tikzpicture}
\end{center}
%[
%    dnode/.style={draw, fill=black, minimum width=1.25cm, minimum height=1.25cm}
%    enode/.style={draw, fill=white, minimum width=1.25cm, minimum height=1.25cm}]
%    
%\matrix[row sep=-\pgflinewidth, column sep=-\pgflinewidth,] (A) {
%    \node[dnode]{$D_1$}; & \node[enode]{$D_2$}; & \node{\dots}; & \node[enode]{$D_n$};\\
%    \node[dnode]{$F_1$};\\
%     & \node[dnode]{$F_2$};\\
%    & & \node{$\ddots$};\\
%    & & &\node[dnode]{$F_n$};\\    
%    };

%\begin{equation}
%\begin{bNiceMatrix}[margin][cell-space-limits=6pt]
% \Block[fill=black]{1-1}{2-2} &  \Block[fill=gray]{1-3}{3-3}  
%  a_{21} & a_{22} & \cdots & a_{2n} & b_2\\
%  \vdots & \vdots  & \ddots & \vdots & \vdots\\
%  a_{n1} & a_{n2}  & \cdots & a_{nn} & b_n
%%\CodeAfter \tikz \draw (1-|5) |- (2-|last) ; 
%\end{bNiceMatrix}
%\end{equation}   

In the first model, the diagonal blocks are complete matrices and the off-diagonal blocks are
diagonal.
In the second model, the diagonal blocks are diagonal matrices, but the off-diagonal blocks are complete.

More specifically, let $m$ and $n$ be two positive integers. In the first model, we consider block matrices of size $mn\times mn$ in which
$m$ blocks of size $n\times n$ are on the diagonal, and the remaining off-diagonal blocks are diagonal. 

Let $s_i^2$ for $1 \le i \le mn$, $\sigma_i^2$ for 
$1 \le i \le m$, and $\tau_i^2$ for $1 \le i \le n$ be positive numbers. Assume that $\mathbf{M} \in \R^{mn\times mn}$ has 
the following decomposition
\begin{equation}
\label{eqn:DSF}
\mathbf{M}  = \mathbf{D} + \mathbf{S} + \mathbf{F}
\end{equation}
where $\mathbf{D}$ is a diagonal matrix with entries
\begin{equation}
%\label{eqn:DMPrimal}
\mathbf{D}_{i,i} = s_i^2
\end{equation}
and $\mathbf{S}$ is a block diagonal matrix given by
\begin{equation}
\mathbf{S}  = 
\begin{bNiceArray}{@{\hspace*{10pt}}cccc@{\hspace*{10pt}}}[cell-space-limits=2pt]
  \mathbf{S}_{1} & \mathbf{0}  & \cdots &\mathbf{0} \\[0.4ex]
  \mathbf{0} & \mathbf{S}_{2}  & \cdots &\mathbf{0} \\[0.4ex]
  %\mathbf{0} & \mathbf{0} & \mathbf{S}_{3} & \cdots &\mathbf{0} \\[0.5ex]
  \vdots    &\vdots              & \ddots &\vdots    \\[0.4ex]
  \mathbf{0} & \mathbf{0}  &\cdots &\mathbf{S}_{m}
\end{bNiceArray}
\end{equation}
where $\mathbf{S}_{1}, \mathbf{S}_{2}, \ldots, \mathbf{S}_{m} \in \R^{n\times n}$ with
\begin{equation}
%\label{eqn:IMPrimal}
\mathbf{S}_i = \sigma_i^2\mathbf{J}_{n}
\end{equation}
Finally, $\mathbf{F}$ is a block matrix with $m^2$ identical blocks $\mathbf{E}$ as
\vspace*{4ex}
\begin{equation}
\mathbf{F}  = 
\begin{bNiceArray}{@{\hspace*{10pt}}cccc@{\hspace*{10pt}}}[cell-space-limits=2pt]
  \mathbf{E} & \mathbf{E}  & \cdots &\mathbf{E} \\[0.4ex]
  \mathbf{E} & \mathbf{E}  & \cdots &\mathbf{E} \\[0.4ex]
 % \mathbf{E} & \mathbf{E} & \mathbf{E} & \cdots &\mathbf{E} \\[0.5ex]
  \vdots    &\vdots                & \ddots &\vdots    \\[0.4ex]
  \mathbf{E} & \mathbf{E} &\cdots &\mathbf{E}
 \CodeAfter
  \OverBrace[shorten,yshift=3pt]{1-1}{4-4}{\text{$m$ blocks}}
\end{bNiceArray}
\end{equation}
where each block $\mathbf{E} \in \R^{n\times n}$ is a diagonal matrix with entries
\begin{equation}
\mathbf{E}_{i,i} = \tau_i^2
\end{equation}
Alternatively, $\mathbf{F}$ can be represented as the Kronecker product of $\mathbf{E}$ 
and $\mathbf{J}_{m}$, i.e., $\mathbf{F} = \mathbf{E}\otimes \mathbf{J}_{m}$ \cite{steeb2011matrix}.

In Section~\ref{sec:NFGEquiv}, we will demonstrate that $\mathbf{M}$ can be viewed as the precision matrix of a Gaussian graphical
model defined on a complete bipartite graph. 
We will then present a second Gaussian graphical model, also defined on a complete bipartite graph, 
whose precision matrix is $(m+n)\times(m+n)$ and its determinant is equal $|\mathbf{M}|$ up to scale 
(i.e., the models are equivalent). The equivalence of 
the models follows from the Normal Factor Graph Duality Theorem~\cite{AY:2011}.

\section{The models}
\label{sec:NFGEquiv}

Let $\K_{m,n} = (\V ,\E)$, where 
the vertex set $\V$ is partitioned into $\U$ and $\W$. We assume that $|\U| = m$ and $|\W| = n$.

In the first model, $m$ zero-sum 
indicator factors~(\ref{eqn:Definition2}) are placed at the vertices of $\U$, while $n$ zero-sum 
indicator factors sit at the vertices of $\W$, as illustrated in Fig.~\ref{fig:Dualfactors}.

The labeling of the edges is arbitrary.\footnote{E.g., complete bipartite graphs are \emph{graceful}, i.e., we can 
assign distinct positive integers to the nodes in such a way that the edges are labeled with the absolute differences 
between node values. For more details see~\cite{golomb1972}.} In our adopted labeling, the edges incident
to $u_1, u_2, \ldots, u_m \in \U$ are sequentially labeled with 
the integers $1,2, \ldots, mn$. 

We suppose that $m$ factors $\{g_i(y_i)\}_{i=1}^m$ and $n$ factors $\{h_i(z_i)\}_{i=1}^n$ are attached 
to the vertices of $\U$  
and $\W$, respectively. Additionally, a factor $f_{e}(x_{e})$ is placed on each edge of $\K_{m,n}$. All factors are zero-mean 
univariate Gaussian. We further assume that zero-sum indicator factors sit at the vertices of the model.

The edges of the graph represent random variables $\X = \{X_e, e \in \E\}$.
Thus $Y_1, Y_2, \ldots, Y_m$ and $Z_1, Z_2, \ldots, Z_n$ are linear combinations 
of $\X$, e.g.,
\begin{equation}
Y_1  + X_1 + X_2 + \ldots + X_n = 0
\end{equation}
and
\begin{equation}
Z_n  + X_n + X_{2n} + \ldots + X_{mn} = 0
\end{equation}
as shown in Fig.~\ref{fig:Dualfactors}.

The PDF associated with the model is therefore only a function of $\x$ and can be written as 
\begin{equation}
\label{eqn:ModelPrimalPDF1}
\pi(\x)  \propto \prod_{e \in \E} f_{e}(x_{e})\prod_{i \in [m]} g_i(y_i)\prod_{i \in [n]} h_i(z_i)  
\end{equation}
where 
\begin{equation}
\label{eqn:fkmn}
f_{e}(x_{e})  =  \frac{s_{e}}{\sqrt{2\pi}}\textrm{exp}\Big(-\frac{s^2_{e}x_e^2}{2}\Big), \quad  e \in \E 
\end{equation}
and
\begin{equation}
\label{eqn:gkmn}
g_{i}(y_{i})  =  
\frac{\sigma_{i}}{\sqrt{2\pi}}\textrm{exp}\big(-\frac{\sigma^2_{i}y_{i}^2}{2}\big), \quad 1 \le i \le m
\end{equation}
and
\begin{equation}
\label{eqn:hkmn}
h_{i}(z_{i})  =  
\frac{\tau_{i}}{\sqrt{2\pi}}\textrm{exp}\big(-\frac{\tau^2_{i}z_{i}^2}{2}\big), \quad 1 \le i \le n
\end{equation}
Here, $s_e^2$ for $e \in \E$, $\sigma_i^2$ for 
$1 \le i \le m$, and $\tau_i^2$ for $1 \le i \le n$ denote the precisions (i.e., the inverse variances).

The PDF $\pi(\x)$ can be expressed as the PDF of a multivariate Gaussian distribution as in (\ref{eqn:GaussPDF})
\begin{equation}
\label{eqn:ModelPrimalPDF2}
\pi(\x)   =  \frac{1}{|2\pi \pmb{\Sigma}_\pi|^{1/2}}\textrm{exp}\big(-\frac{ \ 1\ }{ \ 2\ }
\x_{\phantom{1}}^\intercal
\pmb{\Sigma}_\pi^{-1} \x\,\big)
\end{equation}
where $\pmb{\Sigma}_\pi^{-1} \in \R^{mn\times mn}$ is the precision matrix. It can be easily established that
\begin{IEEEeqnarray}{rCl}
\pmb{\Sigma}_\pi^{-1} & = & \mathbf{D}+\mathbf{S}+\mathbf{F} \\
                                      & = & \mathbf{M}
\end{IEEEeqnarray}
from (\ref{eqn:DSF}). Therefore $\pmb{\Sigma}_\pi^{-1}$ has the desired structure. This result also proves that $\mathbf{M}$ is positive-definite.

The normal factor graph of the model is shown in Fig.~\ref{fig:Dualfactors}. 
The boxes labeled ``$+$" represent zero-sum indicator factors~(\ref{eqn:Definition2}) and the boxes 
labeled ``$=$" represent equality indicator factors~(\ref{eqn:Definition1}). The small empty boxes 
represent factors~(\ref{eqn:fkmn}), (\ref{eqn:gkmn}), and (\ref{eqn:hkmn}).

\subsection{Example}
\label{sec:ExPi}

Let $m = n = 3$. Set $s_i^2 = 2$ for $1 \le i \le 9$ and $\sigma_i^2 = \tau_i^2 = 1$ for $1 \le i \le 3$. In this 
example $\pmb{\Sigma}_\pi^{-1} \in \R^{9\times 9}$ is as in
\begin{equation}
\label{eqn:Amatrix}
\pmb{\Sigma}_\pi^{-1} = 
\begin{bNiceArray}{@{\hspace*{10pt}}ccccccccc@{\hspace*{10pt}}}[cell-space-limits=2pt]
 4 & 1 & 1 & 1 & 0 & 0 & 1 & 0 & 0\\[0.2ex]
 1 & 4 & 1 & 0 & 1 & 0 & 0 & 1 & 0\\[0.2ex]
 1 & 1 & 4 & 0 & 0 & 1 & 0 & 0 & 1\\[0.2ex]
 1 & 0 & 0 & 4 & 1 & 1 & 1 & 0 & 0\\[0.2ex]
 0 & 1 & 0 & 1 & 4 & 1 & 0 & 1 & 0\\[0.2ex]
 0 & 0 & 1 & 1 & 1 & 4 & 0 & 0 & 1\\[0.2ex]
 1 & 0 & 0 & 1 & 0 & 0 & 4 & 1 & 1\\[0.2ex]
 0 & 1 & 0 & 0 & 1 & 0 & 1 & 4 & 1\\[0.2ex]
 0 & 0 & 1 & 0 & 0 & 1 & 1 & 1 & 4
\end{bNiceArray} 
\end{equation}

\noindent
with $\mathbf{D} = 2\mathbf{I}_9$, $\mathbf{S}_i = \mathbf{J}_3$, and $\mathbf{E} = \mathbf{I}_3$.

\begin{figure}[t]
  \centering
  \begin{tikzpicture}[scale=0.22]
   \linethickness{1.7mm}
\node[draw, line width=0.7, rectangle, label=left:{$\delta_{+}$}] at (0,0) (0) {$+$};
%\node[draw, line width=0.7, rectangle, label=below:{$\delta_{+}$}, label=left:{$4$}] at (0,0) (0) {$+$};
\node[draw, line width=0.7, rectangle, label=left:{$\delta_{+}$}] at (8,0) (1) {$+$};
\node[draw, line width=0.7, rectangle, label=left:{$\cdots\cdots\quad\,\delta_{+}$}] at (20,0) (2) {$+$};
\node[draw, line width=0.7, rectangle, label=left:{$\delta_{+}$}] at (0,12) (3) {$+$};
\node[draw, line width=0.7, rectangle, label=left:{$\delta_{+}$}] at (8,12) (4) {$+$};
\node[draw, line width=0.7, rectangle, label=left:{$\cdots\cdots\quad\,\delta_{+}$}] at (20,12) (5) {$+$};
\node[draw, line width=0.7, rectangle, label=right:{$\delta_{=}$}] at (0,6) (6) {$=\!\!\!\!\textcolor{white}{l}$};
%\node[draw, line width=0.7, rectangle, label=above:{$\mathcal{F}h$}] at (8,12) (5) {\textcolor{white}{{\LARGE A}}};
%\node[draw, line width=0.7, rectangle, label=left:{$\mathcal{F}h$}] at (0,6) (3) {\textcolor{white}{{\LARGE A}}};
%\node[draw, line width=0.7, rectangle, label=right:{$\mathcal{F}h$}] at (16,6) (7) {\textcolor{white}{{\LARGE A}}};
%\node[draw, line width=0.7, rotate=-52, rectangle, label=right:{$\mathcal{F}h$}] at (10.5,7.9) (8) {\textcolor{white}{{\LARGE A}}};
%\node[draw, line width=0.7, rotate=52, rectangle, label=left:{$\mathcal{F}h$}] at (5.4,7.9) (9) {\textcolor{white}{{\LARGE A}}};
%
\node[draw, line width=0.7, rectangle, label=left:{$h_{1}$}] at (0,-3.75) (10) {\textcolor{white}{{\tiny A}}};
\node[draw, line width=0.7, rectangle, label=left:{$h_{2}$}] at (8,-3.75) (11) {\textcolor{white}{{\tiny A}}};
\node[draw, line width=0.7, rectangle, label=left:{$h_{n}$}] at (20,-3.75) (12) {\textcolor{white}{{\tiny A}}};
\node[draw, line width=0.7, rectangle, label=left:{$g_{1}$}] at (0,15.75) (13) {\textcolor{white}{{\tiny A}}};
\node[draw, line width=0.7, rectangle, label=left:{$g_{2}$}] at (8,15.75) (14) {\textcolor{white}{{\tiny A}}};
\node[draw, line width=0.7, rectangle, label=left:{$g_{m}$}] at (20,15.75) (15) {\textcolor{white}{{\tiny A}}};
\node[draw, line width=0.7, rectangle, label=above:{$f_{1}$}] at (-5.5,6) (16) {\textcolor{white}{{\tiny A}}};
%\node[draw, line width=0.7, rectangle, label=right:{$\mathcal{F}g_{3}$}] at (19.75,-2.75) (11) {\textcolor{white}{{\tiny A}}};
%\node[draw, line width=0.7, rectangle, label=right:{$\mathcal{F}g_{2}$}] at (19.75,14.75) (12) {\textcolor{white}{{\tiny A}}};
%\node[draw, line width=0.7, rectangle, label=left:{$\mathcal{F}g_{1}$}] at (-3.75,14.75) (13) {\textcolor{white}{{\tiny A}}};
%
\draw[line width=0.7] (0) -- (10);
\draw[line width=0.7] (1) -- (11);
\draw[line width=0.7] (2) -- (12);
\draw[line width=0.7] (3) -- (13);
\draw[line width=0.7] (4) -- (14);
\draw[line width=0.7] (5) -- (15);
\draw[line width=0.7] (0) -- (6);
\draw[line width=0.7] (6) -- (3);
\draw[line width=0.7] (3) -- (1);
\draw[line width=0.7] (3) -- (2);
\draw[line width=0.7] (0) -- (4);
\draw[line width=0.7] (4) -- (1);
\draw[line width=0.7] (4) -- (2);
\draw[line width=0.7] (5) -- (0);
\draw[line width=0.7] (5) -- (1);
\draw[line width=0.7] (5) -- (2);
\draw[line width=0.7] (6) -- (16);
%\draw[line width=0.7] (6) -- (3);
%
\draw (1,13.8) node{$Y_1$};
\draw (9,13.8) node{$Y_2$};
\draw (21.1,13.8) node{$Y_m$};
\draw (0.92,8.7) node{$X_1$};
\draw (2.79,9.4) node{$X_2$};
\draw (3.3,11.1) node{$X_n$};
\draw (10.9,11.0) node{$X_{2n}$};
\draw (21.9,9.9) node{$X_{mn}$};
\draw (-2.7,4.8) node{$X_1$};
\draw (1,-2) node{$Z_1$};
\draw (9,-2) node{$Z_2$};
\draw (21.1,-2) node{$Z_n$};
%\draw (2.55,12.75) node{${\scriptstyle \bigomega_{1}^{(1)}}$};
%\draw (3.55,10.9) node{${\scriptstyle \bigomega_{1}^{(2)}}$};
%\draw (-1.35,9.8) node{${\scriptstyle \bigomega_{1}^{(3)}}$};
%
%\draw (13.5,12.75) node{${\scriptstyle \bigomega_{2}^{(1)}}$};
%\draw (13.15,10.9) node{${\scriptstyle \bigomega_{2}^{(2)}}$};
%\draw (17.35,9.8) node{${\scriptstyle \bigomega_{2}^{(3)}}$};
  \end{tikzpicture}
    \caption{\label{fig:Dualfactors}
The normal factor graph of the Gaussian distribution in~(\ref{eqn:ModelPrimalPDF1}). 
The boxes labeled ``$+$" represent zero-sum indicator factors~(\ref{eqn:Definition2}) and the boxes 
labeled ``$=$" represent equality indicator factors~(\ref{eqn:Definition1}). The small empty boxes 
represent factors~(\ref{eqn:fkmn}), (\ref{eqn:gkmn}), and (\ref{eqn:hkmn}). Certain edges and factors in the model are removed to reduce clutter.}
  \end{figure}
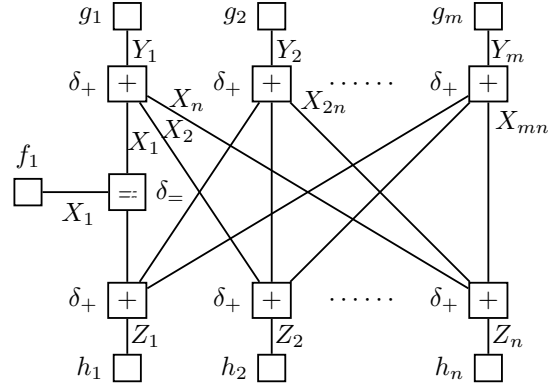

We can construct the dual of Fig.~\ref{fig:Dualfactors}
by adopting the following steps: I) replace each variable, say $x_i$, by the dual variable $\tilde{x}_i$. II) replace 
each factor, say $g_i(\cdot)$, by its Fourier transform $\tilde{g}_i(\cdot)$, which includes replacing equality indicator factors by 
zero-sum indicator factors, and vice versa. III) replace each edge by a 
sign-inverting edge~\cite{Forney:01, Forney:11, Forney:18}.

From (\ref{eqn:ftGaussian}), we can compute the Fourier transform of the local factors~(\ref{eqn:fkmn}), (\ref{eqn:gkmn}), and (\ref{eqn:hkmn}). Indeed
\begin{equation}
\label{eqn:tildefl}
\tilde{f}_{i,j}(\tilde{y}_i, \tilde{z}_j)  =  \textrm{exp}\Big(-\frac{(\tilde{y}_i - \tilde{z}_j)^2}{2s^2_{i,j}}\Big), \quad  
\end{equation}
for $(i,j) \in \E$, and
\begin{equation}
\label{eqn:tildegl}
\tilde{g}_{i}(\tilde{y}_{i})  =  
\textrm{exp}\big(-\frac{\tilde{y}_{i}^2}{2\sigma^2_{i}}\big), \quad 1 \le i \le m
\end{equation}
and
\begin{equation}
\label{eqn:tildehl}
\tilde{h}_{i}(\tilde{z}_{i})  =  
\textrm{exp}\big(-\frac{\tilde{z}_{i}^2}{2\tau^2_{i}}\big), \quad 1 \le i \le n
\end{equation}
where $s_e^2$ for $e \in \E$, $\sigma_i^2$ for 
$1 \le i \le m$, and $\tau_i^2$ for $1 \le i \le n$ denote the corresponding variances.

The PDF is a function of $(\tilde{\y},\tilde{\z})$, and is given by 
\begin{equation}
\label{eqn:globaldualKmn}
\rho(\tilde\y,\tilde\z) \propto \prod_{(i,j) \in \E} \tilde{f}_{i,j}(\tilde{y}_i, \tilde{z}_j)
\prod_{i \in [m]} \tilde{g}_{i}(\tilde{y}_{i})\prod_{i \in [n]} \tilde{h}_{i}(\tilde{z}_{i})
\end{equation}
which can also be written as
\begin{equation}
\label{eqn:ModelDualPDF2}
\rho(\tilde\y,\tilde\z)   =  \frac{1}{|2\pi \pmb{\Sigma}_\rho|^{1/2}}\textrm{exp}\Big(-\frac{ \ 1\ }{ \ 2\ }
\begin{bmatrix}
\tilde{\y} \\
\tilde{\z}
\end{bmatrix}^\intercal
\!\pmb{\Sigma}^{-1}_{\rho}
\begin{bmatrix}
\tilde{\y} \\
\tilde{\z}
\end{bmatrix}
\,\Big)
\end{equation}
where $\pmb{\Sigma}_\rho^{-1} \in \R^{(m+n)\times (m+n)}$
decomposes into 
\begin{equation}
\label{eqn:AplusB}
\pmb{\Sigma}^{-1}_{\rho} = \mathbf{A} + \mathbf{B}
\end{equation}

Here $\mathbf{A}$ is a diagonal matrix with entries
\begin{equation}
%\label{eqn:K2}
\mathbf{A}_{i, i} = \left\{ \begin{array}{ll}
       \dfrac{1}{\sigma_i^2} + \displaystyle\sum_{t=1}^n\cfrac{1}{s_{i,t}^2}, & \text{if $1\le i \le m$} \\[3ex]
       \dfrac{1}{\tau_i^2} + \displaystyle\sum_{t=1}^m\cfrac{1}{s_{t,j}^2}, & \text{if $m+1\le i \le m+n$}
      \end{array}\right.   
\end{equation}
and $\mathbf{B}$ is a block anti-diagonal matrix with the following form\footnote{Indeed, a graph $\G$ is bipartite iff there exists a labeling of $\V$ that gives rise 
to an adjacency matrix with the structure in~(\ref{eqn:BMatrix}), see~\cite[Chapter 2]{asratian1998bipartite}.} 
\begin{equation}
\label{eqn:BMatrix}
\mathbf{B}  = \begin{bNiceArray}{@{\hspace*{8pt}}cc@{\hspace*{8pt}}}[cell-space-limits=2pt]
  \mathbf{0}_{m\times m} & \mathbf{C}_{m\times n}  \\[0.8ex]
  \mathbf{C}_{n\times m}^\intercal & \mathbf{0}_{n\times n}
\end{bNiceArray}
\end{equation}
with $\mathbf{C}_{i,j} = -1/s_{i,j}^2$. 

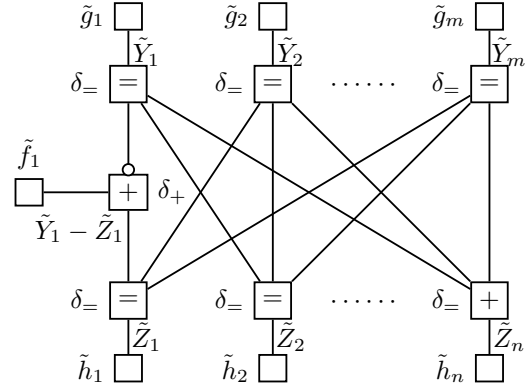
\begin{figure}[t]
  \centering
  \begin{tikzpicture}[scale=0.24, square/.style={regular polygon,regular polygon sides=4, inner sep=1}]
   \linethickness{1.7mm}
\node[draw, line width=0.7, square, label=left:{$\delta_{=}$}] at (0,0) (0) {$=$};
%\node[draw, line width=0.7, rectangle, label=below:{$\delta_{+}$}, label=left:{$4$}] at (0,0) (0) {$+$};
\node[draw, line width=0.7, square, label=left:{$\delta_{=}$}] at (8,0) (1) {$=$};
\node[draw, line width=0.7, square, label=left:{$\cdots\cdots\quad\,\delta_{=}$}] at (20,0) (2) {$+$};
\node[draw, line width=0.7, square, label=left:{$\delta_{=}$}] at (0,12) (3) {$=$};
\node[draw, line width=0.7, square, label=left:{$\delta_{=}$}] at (8,12) (4) {$=$};
\node[draw, line width=0.7, square, label=left:{$\cdots\cdots\quad\,\delta_{=}$}] at (20,12) (5) {$=$};
\node[draw, line width=0.7, rectangle, label=right:{$\delta_{+}$}] at (0,6) (6) {$+$};
%\node[draw, line width=0.7, rectangle, label=above:{$\mathcal{F}h$}] at (8,12) (5) {\textcolor{white}{{\LARGE A}}};
%\node[draw, line width=0.7, rectangle, label=left:{$\mathcal{F}h$}] at (0,6) (3) {\textcolor{white}{{\LARGE A}}};
%\node[draw, line width=0.7, rectangle, label=right:{$\mathcal{F}h$}] at (16,6) (7) {\textcolor{white}{{\LARGE A}}};
%\node[draw, line width=0.7, rotate=-52, rectangle, label=right:{$\mathcal{F}h$}] at (10.5,7.9) (8) {\textcolor{white}{{\LARGE A}}};
%\node[draw, line width=0.7, rotate=52, rectangle, label=left:{$\mathcal{F}h$}] at (5.4,7.9) (9) {\textcolor{white}{{\LARGE A}}};
%
\node[draw, line width=0.7, rectangle, label=left:{$\tilde{h}_{1}$}] at (0,-3.75) (10) {\textcolor{white}{{\tiny A}}};
\node[draw, line width=0.7, rectangle, label=left:{$\tilde{h}_{2}$}] at (8,-3.75) (11) {\textcolor{white}{{\tiny A}}};
\node[draw, line width=0.7, rectangle, label=left:{$\tilde{h}_{n}$}] at (20,-3.75) (12) {\textcolor{white}{{\tiny A}}};
\node[draw, line width=0.7, rectangle, label=left:{$\tilde{g}_{1}$}] at (0,15.75) (13) {\textcolor{white}{{\tiny A}}};
\node[draw, line width=0.7, rectangle, label=left:{$\tilde{g}_{2}$}] at (8,15.75) (14) {\textcolor{white}{{\tiny A}}};
\node[draw, line width=0.7, rectangle, label=left:{$\tilde{g}_{m}$}] at (20,15.75) (15) {\textcolor{white}{{\tiny A}}};
\node[draw, line width=0.7, rectangle, label=above:{$\tilde{f}_{1}$}] at (-5.5,6) (16) {\textcolor{white}{{\tiny A}}};
%\node[draw, line width=0.7, rectangle, label=right:{$\mathcal{F}g_{3}$}] at (19.75,-2.75) (11) {\textcolor{white}{{\tiny A}}};
%\node[draw, line width=0.7, rectangle, label=right:{$\mathcal{F}g_{2}$}] at (19.75,14.75) (12) {\textcolor{white}{{\tiny A}}};
%\node[draw, line width=0.7, rectangle, label=left:{$\mathcal{F}g_{1}$}] at (-3.75,14.75) (13) {\textcolor{white}{{\tiny A}}};
%
\draw[line width=0.7] (0) -- (10);
\draw[line width=0.7] (1) -- (11);
\draw[line width=0.7] (2) -- (12);
\draw[line width=0.7] (3) -- (13);
\draw[line width=0.7] (4) -- (14);
\draw[line width=0.7] (5) -- (15);
\draw[line width=0.7] (0) -- (6);
\draw[line width=0.7] (6) -- (3);
\draw[line width=0.7] (3) -- (1);
\draw[line width=0.7] (3) -- (2);
\draw[line width=0.7] (0) -- (4);
\draw[line width=0.7] (4) -- (1);
\draw[line width=0.7] (4) -- (2);
\draw[line width=0.7] (5) -- (0);
\draw[line width=0.7] (5) -- (1);
\draw[line width=0.7] (5) -- (2);
\draw[line width=0.7] (6) -- (16);
%\draw[line width=0.7] (6) -- (12);
%
\draw (1,13.8) node{$\tilde{Y}_1$};
\draw (9,13.8) node{$\tilde{Y}_2$};
\draw (21.1,13.8) node{$\tilde{Y}_m$};
\draw (-2.7,4.0) node{$\tilde{Y}_1 - \tilde{Z}_1$};
\draw (1,-2) node{$\tilde{Z}_1$};
\draw (9,-2) node{$\tilde{Z}_2$};
\draw (21.1,-2) node{$\tilde{Z}_n$};
\draw[black,fill=white, line width=0.7] (0,7.25) circle (2ex);
  \end{tikzpicture}
    \caption{\label{fig:Primalfactors}
The normal factor graph of the PDF in~(\ref{eqn:globaldualKmn}). 
The boxes labeled ``$+$" are zero-sum indicator factors~(\ref{eqn:Definition2}) and the boxes 
labeled ``$=$" represent equality indicator factors~(\ref{eqn:Definition1}). The small empty boxes 
represent factors~(\ref{eqn:tildefl}), (\ref{eqn:tildegl}), and (\ref{eqn:tildehl}). 
The symbol ``$\circ$" denotes a sign inversion.
Certain edges and factors in the model are removed to reduce clutter.}
%%%%%%%%%%
\end{figure}

Fig.~\ref{fig:Primalfactors} shows the normal factor graph of the PDF in~(\ref{eqn:globaldualKmn}). 
The boxes labeled ``$+$" are equality indicator factors~(\ref{eqn:Definition2}) and the boxes 
labeled ``$=$" represent zero-sum indicator factors~(\ref{eqn:Definition1}). The 
factors~(\ref{eqn:tildegl}), (\ref{eqn:tildehl}), and (\ref{eqn:tildefl}) are represented by the small 
empty boxes. The symbol ``$\circ$" denotes a sign inversion.

%Alternatively, 
%\begin{equation}
%\label{eqn:flz}
%f_{e}(z_{e})  =  \textrm{exp}\big(-\frac{z_e^2}{2s^2_{e}}\big), \quad e \in \E 
%\end{equation}

\subsection{Example}
\label{sec:ExRho}

With the same model parameters as in Example~\ref{sec:ExPi}
\begin{equation}
\label{eqn:ARhomatrix}
\pmb{\Sigma}^{-1}_{\rho} = 
\begin{bNiceArray}{@{\hspace*{6pt}}cccccc@{\hspace*{6pt}}}[cell-space-limits=2pt, columns-width=auto]
 2.5 & 0 & 0 & -0.5 & -0.5 & -0.5 \\[0.3ex]
 0 & 2.5 & 0 & -0.5 & -0.5 & -0.5 \\[0.3ex]
 0 & 0 & 2.5 & -0.5 & -0.5 & -0.5 \\[0.3ex]
 -0.5 & -0.5 & -0.5 & 2.5 & 0 & 0 \\[0.3ex]
-0.5 & -0.5 & -0.5 & 0 & 2.5 & 0 \\[0.3ex]
 -0.5 & -0.5 & -0.5 & 0 & 0 & 2.5 
\end{bNiceArray} 
\end{equation}

\noindent
Here, $\pmb{\Sigma}^{-1}_{\rho} \in \R^{6\times 6}$ decomposes into $2.5\mathbf{I}_6-0.5\mathbf{J}_3$, cf.~(\ref{eqn:AplusB}).

\section{Ratio of Determinants}
\label{sec:Equiv}

According to the Normal Factor Graph Duality Theorem, 
the normalization constant of~(\ref{eqn:ModelPrimalPDF2}) and (\ref{eqn:ModelDualPDF2}) are equal up to scale~\cite{AY:2011}. 
Following the derivation in~\cite{MeMo:2023}, we obtain
\begin{equation}
\label{eqn:SigmaXtoW}
\cfrac{|\pmb{\Sigma}_{\rho}|}{|\pmb{\Sigma}_{\pi}|} = 
\prod_{i \in \left[m\right]} \sigma^2_i \prod_{i \in \left[n\right]} \tau^2_i\prod_{e \in \E} s_e^2
\end{equation}
where, $\pmb{\Sigma}_{\rho} \in \R^{(m+n)\times (m+n)}$ and $\pmb{\Sigma}_{\pi} \in \R^{mn\times mn}$. 

The
determinants are thus equal up to scale. The scale factor only depends on the model parameters and can be easily computed.

In Examples \ref{sec:ExPi} and~\ref{sec:ExRho}, $|\pmb{\Sigma}^{-1}_{\pi}| = 8\times10^4$ and $|\pmb{\Sigma}_{\rho}^{-1}| = 5^4/4$. The ratio
\begin{equation}
\cfrac{|\pmb{\Sigma}_{\rho}|}{|\pmb{\Sigma}_{\pi}|} = 2^9
\end{equation}
is in accordance with (\ref{eqn:SigmaXtoW}).

Different labeling of the edges will give rise to the following transformed precision matrix
\begin{equation}
\label{eqn:transperm}
\pmb{\Sigma}^{-1} \mapsto\, \mathbf{P}\pmb{\Sigma}^{-1}\mathbf{P}_{_{\phantom{1}}}^{\intercal}
\end{equation}
where $\mathbf{P}$ is the corresponding permutation matrix~\cite[Chapter 3]{saad2003iterative}. 
The mapping in~(\ref{eqn:transperm}) 
does not change the value of the determinant. Therefore, our results are applicable to a class of precision matrices 
given by~(\ref{eqn:transperm}).

\section{Exact Determinant of Homogeneous Models}
\label{sec:Exact}

We derive the exact value of $|\pmb{\Sigma}_{\pi}|$ for homogeneous models. For simplicity, we assume 
$m = n$, $s_e^2 = s^2$ for $e \in \E$, and $\sigma^2_i = \tau^2_i = \sigma^2$ 
for $1 \le i \le n$, although the 
exact determinant can be obtained in more general settings.

We first note that from (\ref{eqn:AplusB}), we have
\begin{equation}
\pmb{\Sigma}^{-1}_{\rho} = \mathbf{A} + \mathbf{B}
\end{equation}
where $\mathbf{A}_{ii} = 1/\sigma^2 + n/s^2$, and
\begin{equation}
\label{eqn:IMPrimal}
\mathbf{B}  = \begin{bNiceArray}{@{\hspace*{6pt}}cc@{\hspace*{6pt}}}[cell-space-limits=2pt, columns-width=auto]
  \mathbf{0}_{n\times n} & -\cfrac{1}{s^2}\mathbf{J}_{n}  \\[1.0ex]
  -\cfrac{1}{s^2}\mathbf{J}_{n} & \mathbf{0}_{n\times n}
\end{bNiceArray}
\end{equation}

The eigenvalues of $\mathbf{A}$ are 
$1/\sigma^2 + n/s^2$ with multiplicity $2n$. To compute the eigenvalues of $\mathbf{B}$,
we look at
\begin{equation}
\label{eqn:IMPrimal}
\mathbf{B}^2  =\frac{n}{s^4}\begin{bNiceArray}{@{\hspace*{8pt}}cc@{\hspace*{8pt}}}[cell-space-limits=2pt, columns-width=auto]
 \mathbf{J}_{n}  & \mathbf{0}_{n\times n}  \\[1.0ex]
  \mathbf{0}_{n\times n} & \mathbf{J}_{n}
\end{bNiceArray}
\end{equation}
The eigenvalues of $\mathbf{B}^2$ are zero with multiplicity $2n-2$ and $n^2/s^4$ with multiplicity two. 
Therefore, the eigenvalues of $\mathbf{B}$ are zero with multiplicity $2n-2$, $n/s^2$, and $-n/s^2$.

The eigenvalues of $\pmb{\Sigma}^{-1}_{\rho}$ are the addition of the eigenvalues of $\mathbf{A}$ and $\mathbf{B}$. 
Hence
\begin{IEEEeqnarray}{rCl}
|\pmb{\Sigma}^{-1}_{\rho}| &=& \prod_{i=1}^{2n} \lambda_i\\
& = & \cfrac{1}{\sigma^2}\big(\cfrac{1}{\sigma^2}+\cfrac{2n}{s^2}\big)\big(\cfrac{1}{\sigma^2}+\cfrac{n}{s^2}\big)^{2n-2}
\label{eqn:SigmaDeterm}
\end{IEEEeqnarray}

After applying the scale factor (\ref{eqn:SigmaXtoW}), we obtain
\begin{equation}
|\pmb{\Sigma}^{-1}_{\pi}| = \big(s^2 + 2n\sigma^2\big)(s^2 + n\sigma^2)^{2(n-1)}s^{2(n-1)^2}
\end{equation}
In the limit $n \to \infty$
\begin{equation}
\label{eqn:SigmaDetermLimit}
\lim_{n \to \infty}\frac{\ln|\pmb{\Sigma}_{\pi}|}{n^2} = -2\ln(s)
\end{equation}

\begin{figure}[t!!]
\centering
%\begin{minipage}[t]{0.48\linewidth}
\begin{tikzpicture}
\begin{axis}[
			height = 34.0ex,
			width = 51ex,
			grid = major,
			tick pos=left, 
			xlabel shift = -2 pt,
			ymode = normal,
			xmode=log,
			%restrict y to domain = 0.5:1, %=-10:10
			xminorticks = false,	
		    yminorticks = false,	
		    y tick label style={
        /pgf/number format/.cd,
            fixed,
        /tikz/.cd
    		}, 				
			ytick={-1.2,-1.0,-0.8,-0.6},
			extra y ticks=-0.69314,
			extra y tick labels={$-\ln(2)$},
			%xtick={0, 0.5, 1, 1.5, 2, 2.5, 3},
			%xtick={0.1, 0.7, 1.3, 1.9, 2.5},
			%xtick={4, 8, 16, 32},
			%extra x tick style={grid=none},
			%extra x tick labels={},
		xlabel= $n^2$ ={font=\normalsize},
			xmin = 25,
			xmax = 100000000,
			ymin = -1.2,
			ymax = -0.6,
			%ylabel = $\ln|\pmb{\Sigma}_{\pi}|/n^2$ = {font=\tiny},
			yticklabel style = {font=\tiny,yshift=0.0ex},
            xticklabel style = {font=\tiny,xshift=0.0ex}			
			]

%\addplot table[x=x, y=y] {converge.txt};

\pgfplotstableread{./converge.txt}\mydataone
\pgfplotstableread{./lim.txt}\mydatatwo

		\addplot [
		smooth,
		line width = 0.7 pt,
		 color = black
		]		
		 table[y = Z] from \mydataone; 	 	

		\addplot [
		densely dashed,
		line width = 0.6 pt,
		 color = blue
		]		
		 table[y = Z] from \mydatatwo; 	

\end{axis}
\end{tikzpicture}
%\vspace{-2.0ex}
\caption{\label{fig:SigmaBoundPlot}%
The plots shows $\ln|\pmb{\Sigma}_{\pi}|/n^2$ vs.~$n^2$ for a homogeneous model $\K_{n,n}$ 
with $s^2 = 2$ and $\sigma^2 = 1$. The horizontal blue dashed line shows $\lim_{n \to \infty} \ln|\pmb{\Sigma}_{\pi}|/n^2$, which
is equal to $-\ln(2)$ in this example.}
%\end{minipage}
%\hfill
\end{figure}
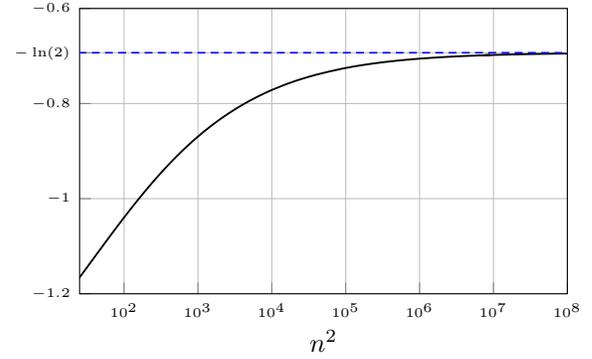

Fig.~\ref{fig:SigmaBoundPlot} shows $\ln|\pmb{\Sigma}_{\pi}|/n^2$ as a function of $n^2$ for $\K_{n,n}$ with 
$s^2 = 2$ and $\sigma^2 = 1$. The blue dashed line shows the limit in (\ref{eqn:SigmaDetermLimit}),
which is equal to $-\ln(2)$ in this example.

\section{A Special Case: $\K_{1,n}$}
\label{sec:NFGK1N}

This section examines the complete bipartite graph $\K_{1,n}$, also known as the \emph{star} graph. 
Since $\K_{1,n}$ is cycle-free, the normalization constants in (\ref{eqn:ModelPrimalPDF2}) and (\ref{eqn:ModelDualPDF2}) can be computed 
via the Gaussian belief propagation algorithm~\cite{Weiss1999correctness}. Consequently, $|\pmb{\Sigma}_{\rho}|$ and $|\pmb{\Sigma}_{\pi}|$ can be
calculated efficiently. 

As in Section~\ref{sec:NFGEquiv}, we assume that $\X_{\phantom{1}}^\intercal = (X_1, X_2, \ldots, X_n)$ are 
represented by the edges of the graph. In the star graph
\begin{equation}
Z + X_1  + X_2 + \ldots + X_{n} = 0
\end{equation}
as illustrated in Fig.~\ref{fig:Dualfactors}. 

The PDF $\pi(\x)$ defined on $\K_{1,n}$ is 
as follows
\begin{equation}
\label{eqn:piforK1n}
\pi(\x)  \propto h(z)\prod_{i \in [n]} f_{i}(x_{i})g_i(x_i)  
\end{equation}
where 
\begin{equation}
f_{i}(x_{i})  =  \frac{1}{\sqrt{2\pi s^i_{e}}}\textrm{exp}\Big(-\frac{s^2_{i}x_i^2}{2}\Big)
\end{equation}
and
\begin{equation}
g_{i}(x_{i})  =  
\frac{1}{\sqrt{2\pi \sigma^2_{i}}}\textrm{exp}\big(-\frac{\sigma^2_{i}x_{i}^2}{2}\big)
\end{equation}
for $1 \le i \le n$, and
\begin{IEEEeqnarray}{rCl}
h(z) & = & \frac{1}{\sqrt{2\pi \tau^2_{i}}}\textrm{exp}\big(-\frac{\tau^2z^2}{2}\big) \\[1ex]
	& = & \frac{1}{\sqrt{2\pi \tau^2_{i}}}\textrm{exp}\Big(-\frac{\tau^2(x_1+x_2+\ldots+x_{n})^2}{2}\Big)
\end{IEEEeqnarray}

The corresponding precision matrix $\pmb{\Sigma}_\pi^{-1} \in \R^{n\times n}$ is 
\begin{equation}
\pmb{\Sigma}_{\pi\,i,j}^{-1}= \left\{ \begin{array}{ll}
        s_{i}^2 + \sigma_i^2 + \tau^2, & \text{if $i=j$} \\[1ex]
       \tau^2, & \text{otherwise.}
      \end{array}\right.   
\end{equation}

In the dual model
\begin{equation}
\label{eqn:rhoforK1n}
\rho(\tilde\y,\tilde z) \propto \tilde{h}(\tilde{z})\prod_{i \in [n]} \tilde{f}_{i}(\tilde{y}_i, \tilde{z})
\tilde{g}_{i}(\tilde{y}_{i})
\end{equation}
where
\begin{equation}
\label{eqn:flk1nd}
\tilde{f}_{i}(\tilde{y}_i, \tilde{z})  =  \textrm{exp}\Big(-\frac{(\tilde{y}_i - \tilde{z})^2}{2s^2_{i}}\Big)
\end{equation}
and
\begin{equation}
\label{eqn:gl}
\tilde{g}_{i}(\tilde{y}_{i})  =  
\textrm{exp}\big(-\frac{\tilde{y}_{i}^2}{2\sigma^2_{i}}\big)
\end{equation}
for $1 \le i \le n$, and
\begin{equation}
\label{eqn:gl}
\tilde{h}(\tilde{z})  =  
\textrm{exp}\big(-\frac{\tilde{z}^2}{2\tau^2}\big)
\end{equation}

The precision matrix $\pmb{\Sigma}^{-1}_{\rho} \in \R^{(n+1)\times (n+1)}$ %associated with $\rho(\tilde\y,\tilde z)$
is a symmetric positive-definite arrowhead matrix\footnote{An arrowhead matrix is a square matrix containing zeros in all entries except for the last row, the last column, and 
the main diagonal~\cite[p. 64]{watkins2010fundamentals}.} with the following structure
\begin{equation}
\label{eqn:ArrowDual}
\pmb{\Sigma}^{-1}_{\rho}  =  \begin{bNiceArray}{@{\hspace*{8pt}}cc@{\hspace*{8pt}}}[cell-space-limits=2pt]
  \mathbf{D} & \mathbf{v}  \\[0.8ex]
  \mathbf{v}_{\phantom{1}}^\intercal & c
\end{bNiceArray}
\end{equation}
Here $\mathbf{D} \in \R^{n\times n}$ is a diagonal matrix with entries
\begin{equation}
\label{eqn:DMPrimal}
\mathbf{D}_{i,i} = \cfrac{1}{\sigma^2_{i}} + \cfrac{1}{s_{i}^2}
\end{equation}
The vector $\mathbf{v} \in \R^{n}$ is given by
\begin{equation}
\label{eqn:AMPrimal}
\mathbf{v}_i =  -\cfrac{1}{s^2_{i}}
\end{equation}
and the scalar $c \in \R$ is
\begin{equation}
\label{eqn:cPrimal}
c = \cfrac{1}{\tau^2} + \sum_{i \in [n]}\cfrac{1}{s_{i}^2}
\end{equation}

%%%%%%%%%%%%%%%%%%%%%%%%%%%%%%%%%%%%%%%%%%%%%%%%%%%%%

For simplicity, we assume that the model is
homogeneous (i.e., $s_i^2 = s$ and $\sigma_i^2 = \sigma$ for $1 \le i \le n$). 

Therefore
\begin{equation}
\pmb{\Sigma}_{\pi}^{-1} = \tau^2\mathbf{J}_{n} + (\sigma^2+s^2)\mathbf{I}_n
\end{equation}
which gives
\begin{equation}
\label{eqn:detSpecialPrimal}
|\pmb{\Sigma}_{\pi}^{-1}| = (\sigma^2+s^2 + n\tau^2)(\sigma^2+s^2)^{n-1}
\end{equation}
After applying the scale factor in~(\ref{eqn:SigmaXtoW}), we obtain
\begin{IEEEeqnarray}{rCl}
|\pmb{\Sigma}_{\rho}^{-1}| & = & \tau^2\sigma^{2n} s^{2n} |\pmb{\Sigma}_{\rho}^{-1}| \\
& = & (\cfrac{1}{\sigma^2} + \cfrac{1}{s^2})^n(\cfrac{1}{\tau^2} + \cfrac{n}{s^2} - \cfrac{n\sigma^2(\sigma^2+s^2)}{s^2})
\label{eqn:detSpecialDual}
\end{IEEEeqnarray}
%$|\mathbf{D}|(c-\mathbf{v}_{\phantom{1}}^\intercal\mathbf{D}^{-1}\mathbf{v})$
The determinant of (\ref{eqn:ArrowDual}) is also derived in~\cite{jakovvcevic2024inverses}.

\subsection{Example}
Let $n = 5$, and set $s_i^2 = 2$ and $\sigma_i^2 = 1$ for $1 \le i \le 4$, and 
$\tau^2 = 1$. With these values
\begin{equation}
\label{eqn:ARhomatrix}
\pmb{\Sigma}^{-1}_{\pi} = 
\begin{bNiceArray}{@{\hspace*{6pt}}ccccc@{\hspace*{6pt}}}[cell-space-limits=2pt, columns-width=auto]
 4 & 1 & 1 & 1 & 1\\[0.3ex]
 1 & 4 & 1 & 1 & 1\\[0.3ex]
 1 & 1 & 4 & 1 & 1 \\[0.3ex]
 1 & 1 & 1 & 4 & 1 \\[0.3ex]
1 & 1 & 1 & 1 & 4
\end{bNiceArray} 
\end{equation}
and
\begin{equation}
\label{eqn:ARhomatrix}
\pmb{\Sigma}^{-1}_{\rho} = 
\begin{bNiceArray}{@{\hspace*{6pt}}cccccc@{\hspace*{6pt}}}[cell-space-limits=2pt, columns-width=auto]
1.5 & 0 & 0 & 0 & 0 & -0.5\\[0.32ex]
0 & 1.5 & 0 & 0 & 0 & -0.5\\[0.32ex]
0 & 0 & 1.5 & 0 & 0 & -0.5 \\[0.32ex]
0 & 0 & 0 & 1.5 & 0 & -0.5\\[0.32ex]
0 & 0 & 0 & 0 & 1.5 & -0.5\\[0.32ex]
-0.5 & -0.5 & -0.5 & -0.5 & -0.5 & 3.5
\end{bNiceArray} 
\end{equation}

\noindent
In this example, $|\pmb{\Sigma}^{-1}_{\pi}| = 648$ and $|\pmb{\Sigma}^{-1}_{\rho}| = 81/4$. The ratio
$|\pmb{\Sigma}_{\rho}|/|\pmb{\Sigma}_{\pi}| =2^5$ is in agreement with (\ref{eqn:SigmaXtoW}).

\section{Conclusion}
\label{sec:Conclusion}

We presented two Gaussian graphical models on complete bipartite graphs, and 
demonstrated that the ratio of the determinants of their precision matrices only 
depends on model parameters.
This equivalence reduces the complexity of computing the exact 
determinant of large precision matrices. Furthermore, it allows the derivation of 
closed-form expressions for the determinants in certain special cases, such as homogeneous models and star graphs.

\balance

\bibliography{mybib}

\end{document}